\documentstyle[epsf,eqsecnum,aps]{revtex}

\def\0#1{\stackrel{(0)}{#1}}
\def\1#1{\stackrel{(1)}{#1}}
\def\Oll{\frac{\nabla_l\nabla^l\Omega}{\Omega}}
\def\Oij{\frac{\nabla_i\nabla_j\Omega}{\Omega}}
\def\OlOl{\frac{\nabla_l\Omega\nabla^l\Omega}{\Omega^2}}
\def\OiOj{\frac{\nabla_i\Omega\nabla_j\Omega}{\Omega^2}}
\def\apm{a_\pm}
\def\aeff{a_{\mbox{\sl eff}}}
\def\3{{}^3\!}
\def\lw{{\hskip -3.5mm}}

\begin{document}
\def\pabnum{{TIT/HEP-324/COSMO-69}\\
	    {KUCP/U-0098}}

\begin{flushright}
\pabnum \\
Jun 8, 1996
\end{flushright}
\begin{center}
{\bf
\vskip 0.25cm
{\Large Inhomogeneity of Spatial Curvature for Inflation }}\\
\vskip 0.8cm

Osamu Iguchi
\footnote{JSPS junior fellow,\quad e-mail: osamu@th.phys.titech.ac.jp},
Hideki Ishihara
\footnote{email: ishihara@th.phys.titech.ac.jp}
and Jiro Soda
\footnote{email: jsoda@phys.ualberta.ca\quad or \quad
jiro@phys.h.kyoto-u.ac.jp} \\ 
	
{ Department of Physics, Tokyo Institute of Technology,
  Meguroku, Tokyo 152, Japan }\\
{ \ddag Department of Physics, University of Alberta, Edmonton, 
  Canada T6G 2J1\\
  and Department of Fundamental Sciences, 
  FIHS, Kyoto University, Kyoto 606, Japan}

\vskip 0.5cm

{\bf abstract}
\vskip 0.15cm

\begin{minipage}[c]{9.5cm}
\small

We study how the initial inhomogeneities of the spatial 
curvature affect the onset of inflation in the closed 
universe. 
We consider a cosmological model which contains a radiation 
and a cosmological constant. 
In order to treat the inhomogeneities in the closed universe, 
we improve the long wavelength approximation such that 
the non-small spatial curvature is tractable in the lowest order. 
Using the improved scheme, we show how large inhomogeneities 
of the spatial curvature prevent the occurrence 
of inflation.

\end{minipage}
\end{center}
\vspace{0.5cm }

\setcounter{equation}{0}
\section{Introduction}

Inflationary universe model\cite{Inf} is the most favorite scenario 
which can explain the large scale homogeneity and isotropy of 
the Universe. 
One of the main questions concerning the inflation 
is whether the occurrence of inflation is generic or not.

There are lots of investigations on this problem in the homogeneous 
model universes. 
For Friedmann-Robertson-Walker universe with an inflaton\cite{review,BIKS}, 
the occurrence of inflation is generic except the closed universe case. 
In the view point of cosmological no-hair conjecture, 
Wald\cite{Wald} showed that initially expanding 
anisotropic Bianchi type universes with a positive 
cosmological constant except the type IX approach de Sitter universe. 
In theses cases, inflation is not completely 
generic because the positive spatial curvature 
can prevent inflation. 

It is obvious that the most appealing feature of the inflationary 
cosmologies 
is smoothing out the initial inhomogeneities of the universe. 
Thus, we should 
clarify the generality of inflation in the initially  
inhomogeneous universe. 
%
%
The effect of inhomogeneities on the inflation is studied by 
numerical simulations 
\cite{GP,Numeric} .
In Ref.\cite{GP}, it is shown that the homogeneity over a few 
initial horizon is needed in order to inflate the universe 
and other works support it.

There are two aspects of the inhomogeneities. One is the 
inhomogeneities of the inflaton, the other is the one in the 
metric. In general these are coupled. 
In this paper we study how the inhomogeneities of the spatial 
geometry affect the occurrence of 
inflation by the use of the long wavelength approximation, 
which reduces the Einstein equations to several sets of 
ordinary differential equations.
%
%


By the use of the long wavelength approximation method 
we can construct 
approximate solutions of Einstein equations which describe 
an inhomogeneous 
universe on scales larger than the local Hubble radius. 
Belinski, Khalatnikov and Lifshitz\cite{BKL} studied the 
general behaviors of 
the universe in the neighborhood of initial singularity by 
this approximation, 
and Tomita\cite{Tomita} focused attentions on the evolutions 
of inhomogeneities of the early universe in this framework.

The long wavelength approximation assumes that 
the spatial derivative of a metric is smaller than 
the time derivative. 
The spatial curvature which consists of the spatial derivative of 
the second order is neglected in the lowest order of this approximation 
and it is taken into account 
perturbatively by the series expansion in the higher order. 
This approximation is called the gradient expansion method 
for the Einstein 
equations. 
For the higher order solutions,  Comer, Deruelle, 
Langlois {\it et al}\cite{Comer}  
developed an iteration scheme of Einstein equations and Salopek,  
Stewart {\it{et al}}\cite{Salopek} developed one of 
the Hamilton-Jacobi equation for general relativity.

%
%
%
Recently, using the gradient expansion method, 
the generality of inflation 
is discussed by Deruelle and Goldwirth\cite{DG}, 
and Nambu and Taruya\cite{NT}. 
As shown by the investigations of the homogeneous models\cite{BIKS,Wald}, 
the positive spatial curvature 
plays an essential role; if the positive curvature dominates over the 
vacuum energy, which drives inflation, 
the universe never enter the inflationary era and recollapse to the 
singularity. 
In the marginal solution between inflationary and recollapsing 
solution, the spatial curvature becomes as large as 
the vacuum energy.
For desired inflation to explain the homogeneity of the universe, 
large vacuum energy is necessary. 
Then we should take account of non-small spatial 
curvature to clarify the condition for the 
occurrence of inflation. 
Readers would suspect that the gradient expansion method is not 
suitable for investigation of the generality of inflation 
because the spatial curvature is assumed to be a small quantity. 

In this paper, we improve the gradient expansion scheme 
so that the non-small constant 
spatial curvature can be tractable in the lowest order.  
By this improved gradient expansion scheme, 
we study how the inhomogeneities of the spatial curvature affect 
the occurrence of inflation in the closed universe. 
The word \lq closed universe\rq\ does not mean the universe 
with spatial surface of closed topology in this paper. 
We investigate the time evolution of the spatial 
region with non-small positive spatial curvature. 

For the first step to investigate the effect of the inhomogeneities 
of the spatial curvature for the onset of inflation 
we consider the model in which the cosmological constant gives 
rise to inflation instead of the scalar field and 
a radiation fluid fills the universe. 

This paper is organized as follows.
In Sec.2, we explain the improved gradient expansion method in 
order to construct the inhomogeneous closed universe.
And we derive the basic equations in the 
first order approximation. 
In Sec.3, we specify the basic equations for our simple model 
and show numerical results for our aim.
Sec.4 is devoted to conclusion and discussions.

\setcounter{equation}{0}
\section{Gradient Expansion from Closed Friedmann Universe}

In this section, we review the usual gradient expansion method 
briefly and modify it 
so that non-vanishing spatial curvature is included in the lowest order. 

\subsection{Brief Review}

We start from a brief review of the usual gradient expansion 
method. 

%
Take a synchronous reference frame where the line element is
\begin{eqnarray*}
   ds^2&=&-dt^2+\gamma_{ij}(t,x^k)dx^idx^j .
\end{eqnarray*}
Throughout this paper Latin letters will denote spatial indices and
Greek letters spacetime indices. 
%
%
The matter is taken to be a perfect fluid  characterized by
energy-momentum tensor 
\[
	T_{\mu\nu}=(\rho + p)u_{\mu}u_{\nu} + p g_{\mu\nu} , 
\]
where $p, \rho $ and $u_{\mu}$ 
are pressure, energy density and four velocity, respectively. 
We assume the equation of state of the fluid is 
$ p/{\rho} \equiv {\Gamma} -1$, where $\Gamma$ is a constant. 
The Einstein equations with a cosmological constant 
$\Lambda $ are  
\begin{eqnarray}
  \3 R_i^j+\frac{1}{2\sqrt{\gamma}}\frac{\partial}{\partial t}
		(\sqrt{\gamma}K_{i}^{j})
  	&=&\frac{2\dot{K}+K_{l}^{m}K_{m}^{l}-4\Lambda}{4(2-3\Gamma
  	-2\Gamma u_l u^l)}[2\Gamma u_i u^j+\delta_i^j(2-\Gamma)] 
  	+\Lambda\delta_i^j ,   
\label{eq:ein1}\\
  \kappa\rho & = &  \frac{2\dot{K}+K_l^m K_m^l-4\Lambda}{2(2-3\Gamma
  	-2\Gamma u_lu^l)}, 
\label{eq:ein2} \\
  \kappa\Gamma\rho u_i 
	& = & -\frac{1}{2\sqrt{1+u_lu^l}} (K_{i;j}^j-K_{, i}) , 
\label{eq:ein3}
\end{eqnarray}
where $\3 R_i^j$ is the Ricci tensor associated with $\gamma_{ij}$, 
$K_{ij} \equiv \dot\gamma_{ij}$ is the extrinsic curvature, 
$\gamma\equiv \det\gamma_{ij}$, 
a dot denotes the derivative with respect to $t$, a semicolon denotes
the covariant derivative with respect to $\gamma_{ij}$ and 
$\kappa \equiv 8\pi G$.

We consider the situation in which all spatial gradients are much 
smaller than time derivatives and assume that the spatial metric is 
expanded as 
\begin{eqnarray}
    \gamma_{ij}(t,x) = \0\gamma_{ij}(t,x) + \1\gamma_{ij}(t,x) + \cdots 
		= \sum_p \stackrel{(p)} \gamma_{ij}(t,x) . 
\label{eq:expand}
\end{eqnarray}   
Here the lowest term $\0\gamma_{ij}$ in Eq.(\ref{eq:expand}) is assumed 
in the quasi isotropic form
\begin{equation}
	\0\gamma_{ij}(t,x) = a^2(t) h_{ij}(x) ,
\label{eq:lowest}
\end{equation}
where $a(t)$ is the scale factor in the lowest order 
which depends on only $t$ 
and $h_{ij}(x)$ is an arbitrary function of only spatial variables. 
The next-order term $\1\gamma_{ij}$ is constructed 
by the symmetric tensors of the 
second order in the spatial gradients of $h_{ij}$ in the form  
\begin{equation}
  \1\gamma_{ij}(t,x) 
	= a^2(t)\left [f_1(t)R_{ij}(h) + g_1(t)R(h)h_{ij} \right] , 
\label{eq:next}
\end{equation}
where $R_{ij}(h)$ is the Ricci tensor associated with $h_{kl}(x)$, 
$R(h)\equiv h^{ij}R_{ij}(h)$, 
and $f_1(t)$ and $g_1(t)$ are functions of time.  
The subsequent general terms $\stackrel{(p)} \gamma_{ij}$ 
in Eq.(\ref{eq:expand}) 
are spatial tensors which 
are of order $2p$ in the spatial gradients of $h_{ij}(x)$ with 
time dependent coefficients. 
Since the spatial inhomogeneities of the spacetime is generated 
by $h_{ij}(x)$, then we call it \lq seed metric\rq. 

Substituting Eq.(\ref{eq:lowest}) into Eqs.(\ref{eq:ein1})-(\ref{eq:ein3}) 
and neglecting the spatial derivatives we get the lowest (zeroth) order 
solution 
\begin{eqnarray}  
  &&a(t) = \left[
	\sinh\left(\frac{3\Gamma}{2}H_0 t\right)\right]^{\frac{2}{3\Gamma}} ,
   		\label{eq:ein012} \\
  &&\0\rho  =  \frac{3H_0^2}{\kappa\sinh^2(\frac{3\Gamma}{2} H_0 t)} ,
                             \label{eq:ein013} \\
  &&\0u_i = 0 , \label{eq:ein014}
\end{eqnarray}
where $H_0 \equiv\sqrt{\Lambda /3}$.  
The evolution of scale factor at zeroth order is the same as 
the one of flat Friedmann-Robertson-Walker universe.
So, we call the solution described by Eqs.(\ref{eq:ein012})-(\ref{eq:ein014}) 
\lq\lq locally flat Friedmann solution\rq\rq .

%
%

At next (first) order, substituting the spatial metric 
$\gamma_{ij} = \0\gamma_{ij} + \1\gamma_{ij}$ given by Eqs.(\ref{eq:lowest}) 
and (\ref{eq:next}) 
into Eq.(\ref{eq:ein1}) and taking 
the spatial derivative terms of the seed metric, 
we get ordinary differential equations for the functions of time 
$f_1(t)$ and $g_1(t)$(see ref.\cite{Comer}). 
Similarly, we can calculate the higher order solutions iteratively. 
The locally flat Friedmann universe, 
Eqs.(\ref{eq:ein012})-(\ref{eq:ein014}), 
is used as a starting point to solve Eq.(\ref{eq:ein1}) in the 
expansion scheme, 
thus we call this scheme the gradient expansion 
from the flat Friedmann universe (GEFF). 
In GEFF, the spatial curvature is treated perturbatively 
as a small quantity in the series expansion.

\subsection{Improved method}

%

In this subsection we improve GEFF such that we can treat non-vanishing 
three curvature which is discarded at lowest order. 
We show that when the seed metric has maximally 
symmetry, the closed or open Friedmann solution is 
derived by summing up all terms in GEFF. 
After that we extend the method such that the inhomogeneities 
around the closed or open Friedmann universe are tractable. 

First, we consider a seed metric in the maxmally symmetric form
with non-vanishing curvature:
\begin{eqnarray*}
    h_{ij}(x) = h^{\pm}_{ij}(x) , 
\end{eqnarray*}   
where the Ricci curvature associated with the metric is 
\begin{eqnarray*}
   R_{ij}(h^{\pm}) = \pm 2 h^\pm_{ij} . 
\end{eqnarray*}   
Since all spatial tensors of rank two derived from $h^\pm_{ij}$ 
are proportional to $h^{\pm}_{ij}$ itself,   
then time dependent part in Eq.(\ref{eq:expand}) is factorized in the form
\begin{eqnarray}
 \gamma_{ij}(t,x)  
	&=& a^2(t)\left [h^\pm_{ij} + f_1(t)R_{ij}(h^{\pm}) 
		+ g_1(t)R(h^{\pm})h^{\pm}_{ij}
		+ f_2(t)R(h^{\pm})R_{ij}(h^{\pm}) + \cdots\right] \nonumber \\
	&=& \apm^2(t) h^\pm_{ij}(x) , 
		\label{eq:friedmann1}
\end{eqnarray}
where 
\begin{equation}
 \apm^2(t) \equiv a^2(t) \Big[1 \pm \{2 f_1(t) + 6 g_1(t)\} 
		+ \{12 f_2(t) +\cdots\} +\cdots \Big]. 
		\label{eq:friedmann2}
\end{equation}

Substituting Eq.(\ref{eq:friedmann1}) into 
Eqs.(\ref{eq:ein1})-(\ref{eq:ein2}) 
we obtain the equation which $\apm$ must satisfy: 
\begin{eqnarray}
 \frac{\ddot \apm}{\apm}
	-\frac{2-3\Gamma}{2}\left(\frac{\dot \apm}{\apm}\right)^2
	\mp \frac{2-3\Gamma}{2\apm^2} &=& \frac{\Gamma}{2}\Lambda , 
\label{eq:F-eq1} \\ 
 \left(\frac{\dot \apm}{\apm}\right)^2
	\pm \frac{1}{\apm^2} &=& \frac{1}{3}(\kappa\rho +\Lambda). 
\label{eq:F-eq2} 
\end{eqnarray}
These are nothing but the equations for the closed or 
open Friedmann universe. 
%
%
Of course, the universe described by the solution satisfying these equations 
is far from the flat Friedmann universe 
but no more than the homogeneous universe.  



Next, we consider the inhomogeneities of the spatial curvature. 
Here we think a simple case where the inhomogeneities are 
expressed by one spatial function. 
We take the seed metric to be conformal to the metric of constant 
curvature,
\begin{equation}
	h_{ij}(x) = \Omega^2(x)h^{\pm}_{ij}(x) , 
\end{equation}
where $\Omega(x)$ is an arbitrary spatial function.

In this case, the Ricci and scalar curvature are given by
\begin{eqnarray}
	R_{ij}(h_{kl}(x)) 
		&=& \pm 2 h^{\pm}_{ij}(x) -\Oij + 2 \OiOj -\Oll h^{\pm}_{ij} 
\label{eq:confRij} \\
	R(h_{kl}(x)) 
		&=&  \frac{1}{\Omega^2}\{\pm 6 - 4 \Oll + 2 \OlOl \} , 
\label{eq:confR}
\end{eqnarray}
where $\nabla_i$ is the covariant derivative with respect 
to $h^{\pm}_{ij}$. 
The spatial metric, therefore, in the expansion form Eq.(\ref{eq:expand}) 
becomes 
\begin{eqnarray}
 \gamma_{ij}(t,x) 
	&=& a^2(t) \left[ h_{ij} + f_1(t)R_{ij}(h) 
		+ g_1(t)R(h)h_{ij}
		+ f_2(t)R(h)R_{ij}(h) + \cdots \right] \nonumber\\
	&=& a^2(t) \Bigg[ \left\{ \Omega^2 \pm (2 f_1(t) + 6g_1(t) )
	   +\Omega^{-2} (12 f_2(t) + \cdots ) + \cdots\right\} h^\pm_{ij}
	   \nonumber\\
	 &&\qquad -\left\{(f_1(t) +4 g_1(t)) +\Omega^{-2}(\pm 14 f_2(t) +\cdots)
		+\cdots\right\} \Oll h^{\pm}_{ij} \nonumber\\
	&&\qquad +\{2g_1(t)+\Omega^{-2}(\pm 4 f_2(t)+\cdots)
		+\cdots\}\OlOl h^{\pm}_{ij} \nonumber\\
	 &&\qquad -\{f_1(t) +\Omega^{-2}(\pm 6 f_2(t)+ \cdots)+ \cdots\} \Oij 
	    \nonumber\\
	&&\qquad  +\{ 2f_1(t) +\Omega^{-2}(\pm 12 f_2(t)+ \cdots )+ \cdots \} \OiOj 
	    \nonumber\\
	&&\qquad + \left\{\Omega^{-2}(4f_2(t) +\cdots)+\cdots \right\}  
		\left(\Oll \right)^2 h^{\pm}_{ij}+\cdots  \Bigg] 
	\label{eq:newexp}
\end{eqnarray}
Here we consider the situation 
that the spatial derivative of $\Omega$ is small. 
From Eq.(\ref{eq:confR}), it means that the derivative of spatial curvature 
is smaller than the value of curvature itself. 

Introducing 
\begin{equation}
	a^2_\pm(t,x) \equiv a^2(t)\left[\Omega^2(x)\pm\{ 2 f_1(t) + 6g_1(t) \}
	   +\Omega^{-2}(x) \{ 12 f_2(t) + \cdots \} + \cdots\right], 
	\label{eq:newSF}
\end{equation}
we rewrite Eq.(\ref{eq:newexp}) as
\begin{eqnarray}
	\gamma_{ij}(t,x)&=& a_\pm^2(t,\Omega)\left[ h^{\pm}_{ij} 
		+ \sum_A F_{(2)}^A(t,\Omega)(\nabla^{(2)}\Omega)^A_{ij} 
		+ \sum_A F_{(4)}^A(t,\Omega)(\nabla^{(4)}\Omega)^A_{ij}
		+ \cdots \right. \nonumber \\
		&&\left. \qquad 
		+ \sum_A F_{(2p)}^A(t,\Omega)(\nabla^{(2p)}\Omega)^A_{ij}
		+ \cdots \right] .
	\label{eq:newexp2}
\end{eqnarray}
Here the notation $(\nabla^{(2p)}\Omega)^A_{ij}$ denotes 
symbolically symmetric 
spatial tensors which contain $2p$ spatial gradients of $\Omega(t,x)$, 
where the suffix $A$ distinguishes the tensor belonging to the same class, 
and $F_{(2p)}^A(t,\Omega)$ is a function of $t$ and $\Omega(x)$ 
which is determined by the Einstein equations later.
This is nothing but a re-summation of the usual gradient expansion. 
%
We restrict ourselves to consider the situation which the series 
expansion Eq.(\ref{eq:newexp2}) converges. 

Substituting Eq.(\ref{eq:newexp2}) into Eqs.(\ref{eq:ein1}) and (\ref{eq:ein2}) 
and neglecting the spatial gradients of $\Omega$ we obtain 
Eqs.(\ref{eq:F-eq1}) and (\ref{eq:F-eq2}) again for $a_\pm(t,x)$ 
for the zeroth order solution 
\begin{eqnarray}
	\0\gamma_{ij}\lw^\pm(t,x) 
		&=& a_\pm^2(t, \Omega(x))h^{\pm}_{ij}(x). 
 \label{eq:gamma-lowest}
\end{eqnarray}
In contrast to Eq.(\ref{eq:friedmann2}), the behavior 
of $a_\pm(t,x)$, defined by Eq.(\ref{eq:newSF}),  depends on $\Omega(x)$. 
The solution of Eqs.(\ref{eq:F-eq1}) and (\ref{eq:F-eq2}) is 
the same form of closed (open) Friedmann universe. 
The new expansion scheme Eq.(\ref{eq:newexp2}) is based on the local 
closed (open) Friedmann universe, 
so we call it
``the gradient expansion from a locally closed (open) Friedmann universe''
(GECF (GEOF)). 

Next, we consider the parts of the second order 
in spatial gradients of $\Omega(x)$ in Eq.(\ref{eq:newexp2}). 
The tensors $(\nabla^{(2)}\Omega)^A_{ij}$ are four types and 
$\1\gamma_{ij}\lw^\pm(t,x)$ is in the form 
\begin{eqnarray}
 \1\gamma_{ij}\lw^\pm(t,x)
	&=& \apm^2(t,x)\left[
 	\frac{1}{3}F(t,\Omega)\Oll h^{\pm}_{ij}
 	+\bar F(t,\Omega)\overline{\Oij}
	\right.\nonumber\\
 	&&\qquad \left. +\frac{1}{3}G(t,\Omega)\OlOl h^{\pm}_{ij}
 	+\bar G(t, \Omega)\overline{\OiOj}\right] , 
 \label{eq:gamma-next}
\end{eqnarray}
where $\overline{\nabla_i\nabla_j\Omega/\Omega}$ and 
$\overline{\nabla_i\Omega\nabla_j\Omega/\Omega^2}$ are 
defined by
\begin{eqnarray*}
 	\overline{\Oij}\equiv \Oij-\frac{1}{3}\Oll h^{\pm}_{ij}, 
	\qquad 
	\overline{\OiOj}\equiv \OiOj - \frac{1}{3} \OlOl h^{\pm}_{ij}. 
\end{eqnarray*}

From Eq.(\ref{eq:ein3}) we obtain
\begin{eqnarray}
 u_{i} &=& \frac{(2-3\Gamma)\left[2\left(\dot\apm'/{\apm}
 	-{\dot\apm\apm'}/{\apm^2}\right)\Omega-\dot{\bar F}\right]}
 	{2\Gamma\left[3 \ddot\apm/\apm-\Lambda\right]}
 	\frac{\nabla_{i}\Omega}{\Omega} ,
 \label{eq:u-next}
\end{eqnarray}
where the prime denotes the differentiation with respect to $\Omega$. 

Substituting $\gamma_{ij}=\0\gamma_{ij}\lw^\pm+\1\gamma_{ij}\lw^\pm$ 
and $u_i$ given by 
Eqs.(\ref{eq:gamma-lowest}), (\ref{eq:gamma-next}) and 
(\ref{eq:u-next}) into Eq.(\ref{eq:ein1}), and 
comparing the coefficients of the second order derivative terms 
of $\Omega$, we obtain 
\begin{eqnarray}
 &&\ddot{F}+3\Gamma\frac{\dot\apm}{\apm}\dot{F} 
	= -\frac{(2-3\Gamma)}{\apm^2}\left[\pm F-\bar F
		+2\frac{\apm'\Omega}{\apm}\right], 
\label{eq:f1} \\
 &&\ddot{\bar F}+3\frac{\dot\apm}{\apm}\dot{\bar F} 
	= \frac{2}{\apm^2}\left[\frac{\apm'\Omega}{\apm} 
		+2(\pm\bar F -\bar F)\right] , 
\label{eq:f2} \\
 &&\ddot{\tilde G}+3\Gamma\frac{\dot\apm}{\apm}\dot{\tilde G} 
 	= -\frac{(4-3\Gamma)(2-3\Gamma)}{2\Gamma \apm
 	\left(3\ddot\apm-\Lambda \apm\right)}
 	\left[\dot{\bar F}-2\left(\frac{\dot\apm'}{\apm}
 	-\frac{\dot\apm\apm'}{\apm^2} \right)\Omega\right]^2 \nonumber\\
	&&\qquad\qquad\qquad -\frac{(2-3\Gamma)}{\apm^2}  
	\left[ \pm G +\bar F - {\bar F}'\Omega 
		\pm2 {\bar F}^2 -{\bar F}^2
	+\Omega^2 \left(2\frac{\apm''}{\apm}-\frac{\apm'^2}{\apm^2}\right)\right], 
\label{eq:g1}  \\
 &&\ddot{\hat G}+3\frac{\dot\apm}{\apm}\dot{\hat G} 
	= \frac{(2-3\Gamma)}{\Gamma \apm\left(3\ddot\apm 
		-\Lambda \apm\right)}\left[\dot{\bar F}
 	-2\left(\frac{\dot\apm'}{\apm}-\frac{\dot\apm\apm'}{\apm^2}
 	\right)\Omega\right]^2 \nonumber \\
	&&\qquad\qquad\qquad + \frac{2}{\apm^2}\left[ \pm 2\bar G 
	+2\bar F -2{\bar F}'\Omega -(\pm 2+\frac{1}{2})\bar F^2 
 	-2\left(\frac{\apm'\Omega}{\apm}\right)^2
	+\frac{\apm''\Omega^2}{\apm}\right] ,  
\label{eq:g2} 
\end{eqnarray}
where we put $\tilde G \equiv G + {\bar F}^2$ and 
$\hat G \equiv \bar G -\frac{1}{2}{\bar F}^2$ for simplicity 
of the equation form. 
%
%
%
%
To solve Eq.(\ref{eq:f1})- Eq.(\ref{eq:g2}), 
we need to know the time evolution of $\bar F'$.
Differentiating Eq.(\ref{eq:f2}) we find that $\bar F'$ must 
satisfy the equation
\begin{eqnarray}
 \ddot{\bar F}'+3\frac{\dot a_\pm}{a_\pm}\dot {\bar F}' 
 	&=&  \frac{2}{a_\pm^2}
 	\left[ \frac{a'_\pm}{a_\pm} + \frac{a''_\pm\Omega}{a_\pm} 
	-3\frac{{a'_\pm}^2\Omega}{a_\pm^2}
	-4\frac{a'_\pm}{a_\pm}(\pm \bar F-\bar F) +2 (\pm \bar F'-\bar F')\right] 
\nonumber\\ 
	&&-3\dot{\bar F}\left[
 	\frac{\dot a'_\pm a_\pm - \dot a_\pm a'_\pm}{a^2_\pm} \right]. 
\label{eq:f2'}
\end{eqnarray}

\setcounter{equation}{0}
\section{
  Inflation in the inhomogeneous universe with a positive curvature
}

In this section, we consider how the inhomogeneities 
of spatial curvature affect on the occurrence of inflation. 
For this purpose we think a simple model, in which the inflation 
is driven by a cosmological constant $\Lambda$. 
We construct the inhomogeneous universe by the gradient expansion from 
the closed Friedmann universe which is introduced in the previous 
section. 
The matter is taken to be a radiation fluid with  
$\Gamma = 4/3$ for simplicity. 

We get the lowest order solution of Eqs.(\ref{eq:F-eq1}) and 
(\ref{eq:F-eq2}) for the system in the form
\begin{eqnarray}
 &&a_+^2(t,\Omega(x))
 	\equiv \frac{1}{2H_0^2}\left[2H_0^2\Omega^2\sinh(2H_0t)
 	+1-\cosh(2H_0t)\right], 
\label{eq:LCF-SF} \\
 &&\0\rho = \frac{3H_0^2\Omega^4}{\kappa a_+^4} .
 \label{eq:LCF-ED}
\end{eqnarray}

Here we see the behavior of the locally closed Friedmann solution 
Eq.(\ref{eq:LCF-SF}). 
The solution has an early phase during which the local scale factor 
behaves as $a_+ \propto t^{1/2}$. 
After the period the behavior of local scale factor depends on the 
local value of $\Omega(x)$. 
The local scale factor $a_+(t,x)$ with $\Omega(x) > \Omega_{cr}$  
approaches de Sitter solution 
and $a_+(t,x)$ with $\Omega(x) < \Omega_{cr}$ recollapses to the singularity, 
where $\Omega_{cr}\equiv \sqrt{1/(2H^2_0)}$. 
The local scale factor $a_+(t,x)$ with the critical value of  
$\Omega = \Omega_{cr}$ approaches a static solution
\begin{equation}
	a_+^2 =\frac{3}{2\Lambda}, \quad \0\rho = \frac{\Lambda}{\kappa}. 
\end{equation}
In the early stage $t<<H_0^{-1}$, the local scale factor is proportional 
to $\Omega(x) t^{1/2}$ then the spatial curvature $\3 R$ 
is proportional to $1/(\Omega^2 t)$. 
The fate of local scale factor is determined  
by the value of $\3 R(t,x) t$ in the lowest order. 
The condition for the occurrence of inflation is 
\begin{equation}
	\frac{1}{\Omega^2} < \frac{1}{\Omega_{cr}^2}. 
\label{eq:critO}
\end{equation}
In other words, it is 
\begin{equation}
	\3 R_{init} < \3 R_{cr}(t) \equiv \frac{3}{\Omega_{cr}^2}\frac{1}{H_0 t}, 
\label{eq:critR}
\end{equation}
where $\3 R_{init}$ is the spatial curvature in the early phase $t<<H_0^{-1}$. 

Next, we consider the first order corrections which are 
proportional to the second order of spatial gradients 
of $\Omega(x)$ in the form 
\begin{eqnarray}
 \1\gamma_{ij}\lw^+(t,x) 
	&=& a_+^2(t,\Omega(x)) \Bigg[\frac{1}{3}F(t, \Omega)\Oll h^+_{ij}
 	+\bar F(t, \Omega)\overline{\Oij}  \nonumber\\
 &&\qquad\qquad\qquad +\frac{1}{3}G(t, \Omega)\OlOl h^+_{ij}
 	+\bar G(t, \Omega)\overline{\OiOj}\Bigg] . 
 \label{eq:rcf12}
\end{eqnarray}


From Eqs.(\ref{eq:f1})-(\ref{eq:f2'}), 
the equations for $F, \bar F, G$ and $\bar G$ reduce to 
\begin{eqnarray}
 \ddot F+4\left(\frac{\dot a_+}{a_+}\right)\dot F
 	&=& \frac{2}{a_+^2}\left[ F-\bar F 
	 +\frac{2\Omega^2\sinh(2H_0t)}{a_+^2}\right],  
\label{f1c}\\
 \ddot{\tilde{G}}+4\left(\frac{\dot a_+}{a_+}\right)\dot{\tilde{G}}
	&=& \frac{2}{a_+^2}\left[ \tilde{G}+\bar F- \bar F'\Omega
	 +\frac{\Omega^2\sinh(2H_0t)}{a_+^2}
	\left(2-\frac{3\Omega^2\sinh(2H_0t)}{a_+^2}\right)\right] ,
\label{g1c}\\
 \ddot{\bar F}+3\left(\frac{\dot a_+}{a_+}\right)\dot{\bar F}
	&=&\frac{2}{a_+^2}\left[ \frac{\Omega^2\sinh(2H_0t)}{a_+^2}
	\right], 
\label{f2c}\\
 \ddot{\hat G} +3 \left(\frac{\dot a_+}{a_+}\right)\dot{\hat G}
	&=&\frac{2}{a_+^2}\left[ 2\hat G+2\bar F-2{\bar F}'\Omega
	-\frac{3\bar F^2}{2}
	+\frac{\Omega^2\sinh(2H_0t)}{a_+^2}
	\left(1-\frac{3\Omega^2\sinh(2H_0t)}{a_+^2}
	\right)\right]\nonumber \\
	&& +\frac{1}{2H_0^{4}\Omega^4 a_+^6}\left[ 
	2\Omega^2(\cosh(2H_0t)-1)-H_0 a_+^4\dot{\bar F}\right]^2 . 
\label{g2c}
\end{eqnarray}
For $\bar F'$ we get
\begin{eqnarray}
 \ddot{\bar F}' +3\frac{\dot a_+}{a_+}\dot{\bar F}' 
	&=& \frac{4\Omega\sinh(2H_0t)}{a_+^4}
 		\left[ 1-\frac{2\Omega^2\sinh(2H_0t)}{a_+^2}\right] 
		-3\dot{\bar F}\frac{\Omega}{H_0 a_+^4}(\cosh(2H_0t)-1) .
\label{f2'c}
\end{eqnarray}

We solve Eqs.(\ref{f1c})-(\ref{f2'c}) by numerical integration. 
These coupled ordinary differential equations have growing and decaying 
solutions. In the course of time, since the growing mode dominates the solution 
we concentrate on the growing solution. 
Then we choose initial conditions for 
$F, \bar F, \bar F', \tilde G, \hat G$ in the form
\begin{eqnarray}
 F=\bar F=\bar F'=\tilde{G}=\hat G=0 .
\label{eq:ic1}
\end{eqnarray}
The growing mode is generated by the source terms in the right hand side of
Eqs.(\ref{f1c})-(\ref{f2'c}). 
For the regularity, 
%
%
asymptotic behaviors of these variables should be 
\begin{eqnarray}
 F = \frac{1}{\Omega^2H_0}t, \quad 
 \bar F=\frac{2}{3\Omega^2H_0}t, \quad
 \bar F'=-\frac{4}{3\Omega^3H_0}t, \quad 
 \tilde G=-\frac{1}{2\Omega^2H_0}t, \quad 
 \hat G=-\frac{4}{3\Omega^2H_0}t.
\label{eq:ic2}
\end{eqnarray}


Time evolution of $F$ and $\tilde G$ is shown in Figs.\ref{fig:F} and 
\ref{fig:G}. 
In the case of $\Omega >\Omega_{cr}$, $F$ and $G$ grow for a while and 
become constant as the universe expands exponentially. 
And in the case of $\Omega < \Omega_{cr}$, they grow and diverge and   
the approximation breaks down. 

We demonstrate a time evolution of the spatial scalar curvature 
given by Eq.(\ref{eq:scalarR}) (see Fig.\ref{fig:R}).
When we fix initial time, there are infinite number of combinations of 
parameters 
$(\Omega, \nabla_l\nabla^l\Omega/\Omega, 
\nabla_l\Omega\nabla_l\Omega/\Omega^2)$ 
which give the same value of $\3 R$. Time evolution of $\3 R$ depends on the 
parameters. Fixing $\Omega$, we vary $\nabla_l\Omega\nabla_l\Omega/\Omega^2$
and $\nabla_l\nabla^l\Omega/\Omega$ keeping $\3 R$ constant. 
When $\Omega$ is fixed, since the evolution of $\3\0R$ depends only on 
$\Omega$, 
the time evolution of $\3 R$ depends on 
$\nabla_l\Omega\nabla_l\Omega/\Omega^2$. 
We see from Fig.\ref{fig:R} that when 
$\nabla_l\Omega\nabla_l\Omega/\Omega^2$ is large $\3 R$ grows rapidly. 
 
By the use of the trace part of the corrections, 
we define an effective local scale factor which includes up to the first 
order by 
\begin{eqnarray}
  \aeff(t,x) 
	&\equiv& \left[\det\Big[\0\gamma_{ij}\lw^+
		+\1\gamma_{ij}\lw^+\Big] \right]^{1/6}
		\nonumber \\
	&=& a_+(t,\Omega)\left[ 
 	1+\frac{1}{6}F(t,\Omega)\Oll
         +\frac{1}{6}\tilde G(t,\Omega)
         \OlOl         \right]. 
\label{eq:ELSF}
\end{eqnarray}
The effective local scale factor $\aeff(t,x)$ describes how the universe 
expands at the point $x$. 

We can follow the evolution of the effective local scale
factor $\aeff(t,x)$ by the evolution of $F(t,\Omega)$ 
and $G(t,\Omega)$. 
As is already seen, at lowest order the behavior of local scale factor 
$a_+(t,\Omega)$ is determined by the value of $\Omega(x)$. 
On the other hand, the spatial derivatives of $\Omega$ 
affect the time evolution of $\aeff(t,x)$ at first order. 
It means that the effective local scale factor evolves under 
the influence of the local value of $\Omega$ and  
the value of it in the neighborhood. 
If a point in the three dimensional parameter space 
$(\Omega, \nabla_l\nabla^l\Omega/\Omega, \nabla_l\Omega\nabla_l\Omega/\Omega^2)$ 
%
is specified we know the time evolution 
of the effective local scale factor $\aeff(t,x)$. 

The effective local expansion rate and acceleration rate are 
given by 
\begin{eqnarray*}
 \frac{\dot \aeff}{\aeff} 
	&=& \frac{\dot a_+}{a_+}+\frac{1}{6}\left[\dot F(t,\Omega)\Oll 
		+\dot{\tilde{G}}(t,\Omega) \OlOl \right] , \\
 \frac{\ddot \aeff}{\aeff} 
 	&=& \frac{\ddot a_+}{a_+}
      	+\frac{1}{6}\left[\ddot F(t,\Omega)
	+2\frac{\dot a_+}{a_+}\dot F(t,\Omega) 
             \right] \Oll
             +\frac{1}{6}\left[ \ddot{\tilde G}(t,\Omega)
             +2\frac{\dot a_+}{a_+}\dot{\tilde G}(t,\Omega)\right]\OlOl .
\end{eqnarray*}

%
When the correction terms which contain the second order derivatives grow to be 
as large as the lowest order term, the expansion scheme breaks down. 
Since the first order terms in Eq.(\ref{eq:newexp2}) contain 
all of the second order spatial derivatives 
then when $(\dot a_+/a_+)^2 > \3\0 R$, GECF reduces to GEFF. 
We, thus, assume the approximation is valid  while  
\begin{eqnarray}
	\left|\frac{\det(\0\gamma_{ij}\lw^+ +\1\gamma_{ij}\lw^+)
		-\det\0\gamma_{ij}\lw^+}{\det\0\gamma_{ij}\lw^+}\right| &<& 0.5 , 
\label{eq:errora} \\
	\left|\frac{\3\!\1{R}}{\max\{(\dot a_+/a_+)^2, \3\!\0{R}\}}\right| &<& 0.5 .
\label{eq:errorR}
\end{eqnarray}

We divide the three dimensional parameter space 
$(\Omega, \nabla_l\nabla^l\Omega/\Omega, \nabla_l\Omega\nabla_l\Omega/\Omega^2)$ 
into two regions: 
inflationary region and recollapsing region (see Figs.\ref{fig:cr49} - 
\ref{fig:cr51}). 
The effective local scale factor with the parameters in the inflationary 
region enters the accelerating expansion phase: 
$\dot \aeff/\aeff > 0$ and $\ddot \aeff/\aeff > 0$. 
On the other hand, the effective local scale factor with the parameters 
in the recollapsing region enters recollapsing phase: 
$\dot \aeff/\aeff < 0$ and $\ddot \aeff/\aeff < 0$. 
Between these two regions, there emerges a fuzzy region where 
error becomes large and Eq.(\ref{eq:errora}) or Eq.(\ref{eq:errorR}) 
does not hold before the effective local scale factor enters the inflationary 
or recollapsing phase.


\section{Conclusion and Discussions}

We studied how the inhomogeneities of  spatial curvature 
affect the occurrence of inflation in the inhomogeneous 
universe which is constructed by the gradient expansion 
from the locally closed Friedmann universe, in which 
the inhomogeneities are described by a function $\Omega(x)$ 
and its spatial gradients. 
For the first step, we considered a simple model universe 
with a cosmological constant and radiation fluid. 

In this paper, we expanded the spatial metric up to the second order 
spatial derivatives  $\nabla_i\nabla_j\Omega/\Omega$ and 
$\nabla_i\Omega\nabla_j\Omega/\Omega^2$. 
We solved the ordinary differential equations for the coefficient 
functions of these derivative terms numerically and evolved the 
inhomogeneous spatial metric\cite{TD}. 
We assumed that the 
inhomogeneities were generated by one function $\Omega(x)$ 
but the universe needed not to have any symmetries.


How the inhomogeneities described by $\nabla_l\nabla^l\Omega/\Omega$ 
and $\nabla_{l}\Omega\nabla^{l}\Omega/\Omega^2$ affect the onset of inflation 
is shown in Figs.\ref{fig:cr49} - \ref{fig:cr51}.
The term $\nabla_{l}\Omega\nabla^{l}\Omega/\Omega^2$ tends to prevent 
inflation and positive (negative) $\nabla_l\nabla^l\Omega/\Omega$ helps 
the universe to inflate (collapse). 
This is related to the fact that the spatial curvature grows rapidly 
when $\nabla_{l}\Omega\nabla^{l}\Omega/\Omega^2$ is large.

From Figs.\ref{fig:cr49} - \ref{fig:cr51}, 
we get the criterion for the occurrence of inflation in the vicinity of 
$\Omega/\Omega_{cr}-1 
=\nabla_l\nabla^l\Omega/\Omega=\nabla_l\Omega\nabla^l\Omega/\Omega^2=0$ 
in a simple time independent form: 
\begin{equation}
	\frac{1}{\Omega^2} \left[1+\alpha\left\{-\Oll + \beta\OlOl\right\} \right] 
	< \frac{1}{\Omega_{cr}^2}, 
\label{eq:critO1}
\end{equation}
where $\alpha\sim 5/4$ and $\beta\sim 6$. 
If the condition Eq.(\ref{eq:critO1}) holds, the effective local scale factor 
enters the inflationary phase. 
By the use of Eq.(\ref{eq:scalarR}) in the limit $t\to 0$, 
the criterion (\ref{eq:critO1}) is translated into 
\begin{equation}
 \3 R_{init}
	+\tilde\alpha
	\left\{	\frac{
	(\3 R_{init})_{;l}^{~;l}}{\3 R_{init}}
	+\tilde\beta\frac{
	(\3 R_{init})_{;l}
	(\3 R_{init})^{;l}}{\3 R_{init}^2} \right\}
	< \3 R_{cr}(t) , 
\label{eq:critR1}
\end{equation}
where $\tilde\alpha\sim 9/5$ and $\tilde\beta\sim 4$ and $ \3 R_{cr}(t)$ is 
given in Eq.(\ref{eq:critR}). 
Equations (\ref{eq:critO1}) and (\ref{eq:critR1}) correspond to 
Eqs.(\ref{eq:critO}) and (\ref{eq:critR}).  
Equation(\ref{eq:critR1}) means that the occurrence of inflation 
at a spatial point is determined by $\3 R_{init}$ and its gradients. 
The spatial curvature at an early time $t$ should be less than 
the critical curvature value $\3 R_{cr}(t)$ 
over a region which has several 
times size of the local curvature radius $(\3 R_{init}(t))^{-1/2}$
for the onset of inflation in the future, 
where $\3 R_{cr}(t)$ is the marginal value for inflation of the 
closed Friedmann universe. 


If the homogeneous universe enters the inflationary phase, 
it never recollapse. However, for the inhomogeneous universe 
even if the local scale factor enters the inflationary phase,
it would be possible to recollapse. It may be also possible that 
the recollapsing local scale factor re-expands again in the inhomogeneous 
universe. 
In order to observe the final fate of the inhomogeneous universe, 
we should sum up all order terms of the series expansion. 
It is actually very difficult. 
In this paper we investigated the behavior 
of solutions up to the first order of the expansion 
while the criteria of the validity of the approximation 
Eqs.(\ref{eq:errora}) and (\ref{eq:errorR}) hold. 
For the purpose to know how the inhomogeneities affect the onset of 
inflation, it is worth clarifying the behavior of the approximate 
solutions of the order. 

In the more realistic model of the inflationary universe, a certain 
field variable, 
e.g. a scalar field, drives inflation. In this case, not only the 
spatial curvature but also the field 
has inhomogeneities. 
We will discuss this case elsewhere. 

\vskip 1cm
\noindent
{\Large\bf Acknowledgements}
\vskip 5mm
We would like to thank Professor A.Hosoya for continuous encouragement.
One of authors ( J.S. ) thanks Dr. Salopek for his hospitality. 
This work was supported in part by the Grant-in-Aid for Scientific Research 
No.06740222 and No.5149. 

\setcounter{equation}{0}
\begin{appendix}
\section{Rule of the order of the spatial tensor}

When we calculate the first order approximation, we 
determine the order of spatial tensor which contains the sixth order 
terms in the spatial gradient at least because they appear 
in the spatial curvature $\1{\3R_i^{~j}}$.
Here we explain the rule we adopted when we calculate the first order 
approximation.

For scalar:
$(\nabla^{(2)}\Omega)$ and $(\nabla^{(4)}\Omega)$ are the scalar of 
the second order terms in the spatial gradient and the fourth order 
terms, respectively.
$(\nabla^{(2)}\Omega)$ consists of two terms, 
\begin{eqnarray*}
 \frac{\nabla_l\nabla^l\Omega}{\Omega}, \quad
 \frac{\nabla_l\Omega\nabla^l\Omega}{\Omega^2}.
\end{eqnarray*}
$(\nabla^{(4)}\Omega)$ is classified into the following three types,
\begin{eqnarray*}
\nabla_l\nabla^l(\nabla^{(2)}\Omega),\quad
\nabla_l(\nabla^{(2)}\Omega)\nabla^l\Omega,\quad 
(\nabla^{(2)}\Omega)(\nabla^{(2)}\Omega). 
\end{eqnarray*}

For spatial tensor:
We adopt the tensors $(\nabla^{(4)}\Omega)_{ij}$ consist of 
the following terms, 
\begin{eqnarray*}
 \frac{\nabla_i\nabla_j\nabla_l\nabla^l\Omega}{\Omega}, \\
 \frac{\nabla_i\nabla_j\Omega\nabla_l\nabla^l\Omega}{\Omega^2},\quad 
 \frac{\nabla_i\Omega\nabla_j\nabla_l\nabla^l\Omega}{\Omega^2}, \quad
 \frac{\nabla_l\nabla_i\nabla_j\Omega\nabla^l\Omega}{\Omega^2}, \quad
 \frac{\nabla_i\nabla_j(\nabla_l\Omega\nabla^l\Omega)}{\Omega^2}, \\
 \frac{\nabla_i\Omega\nabla_j\Omega\nabla_l\nabla^l\Omega}{\Omega^3},\quad  
 \frac{\nabla_i\nabla_j\Omega\nabla_l\Omega\nabla^l\Omega}{\Omega^3},\quad 
 \frac{\nabla_i\Omega\nabla_j(\nabla_l\Omega\nabla^l\Omega)}{\Omega^3},\\
 \frac{\nabla_i\Omega\nabla_j\Omega\nabla_l\Omega\nabla^l\Omega}{\Omega^4},
\end{eqnarray*}
because the trace of these terms belong to the above scalar category.

The sixth order terms in the spatial gradient appear in the spatial 
curvature $\1{\3R_i^{~j}}$ are transformed into the following 
terms by use of the commutation relation,
\begin{eqnarray*}
 \frac{\nabla_l\nabla_m\nabla_i\nabla_j\Omega\nabla^l\nabla^m\Omega}{\Omega^2},
 \quad 
 \frac{\nabla_l\nabla_i\nabla_j\Omega\nabla^l\nabla_m\nabla^m\Omega}{\Omega^2},
 \\
 \frac{\nabla_i\nabla_j(\nabla_l\nabla_m\nabla^m\Omega\nabla^l\Omega)}
 {\Omega^2}, \quad
 \frac{\nabla_i\nabla_j(\nabla_l\nabla^l(\nabla_m\Omega\nabla^m\Omega))}
 {\Omega^2}.
\end{eqnarray*}
We treat them as higher order terms. 
They don't affect the calculation in the first order approximation 
because their trace contain the spatial derivative of $(\nabla^{(2)}\Omega)$.

\section{Expression for the spatial curvature in the first order approximation}

In the gradient expansion from the locally closed or open Friedmann universe, 
we expand the spatial metric up to the second order of the spatial 
gradients in the form
\begin{equation}
 \gamma_{ij} = \0\gamma_{ij}\lw^\pm +\1\gamma_{ij}\lw^\pm , 
\end{equation}
where
\begin{eqnarray*}
 \0\gamma_{ij}\lw^\pm &=& a_\pm^2(t,\Omega(x)) h^\pm_{ij}, \\
 \1\gamma_{ij}\lw^\pm &=& a_\pm^2(t,\Omega(x))\left[
 	\frac{1}{3}F(t,\Omega)\Oll h^\pm _{ij}
 	+\bar F (t,\Omega)\overline{\frac{\nabla_i\nabla_j\Omega}{\Omega}}
 	\right. 	\nonumber\\ 	
	&& +\frac{1}{3}G(t,\Omega)
 	\OlOl h^\pm _{ij}
 	\left.+\bar G (t,\Omega)\overline{\frac{\nabla_i\Omega\nabla_j\Omega}
 	{\Omega^2}}\right]. \\
\end{eqnarray*}
From the inverse of ${\gamma_{ij}}$, we get 
\begin{eqnarray}
 	\1\gamma_\pm\lw^{ij} 
 	&=& -\frac{1}{a_\pm^2}\left[ \frac{1}{3}F\Oll h_\pm^{ij}
 		+\bar F 
		\overline{\frac{\nabla^i\nabla^j\Omega}{\Omega}} \right.\nonumber\\
 	&&+\frac{1}{3}(G+2{\bar F}^2) \OlOl h_\pm^{ij}
	+\left.(\bar G-\bar F^2)
		\overline{\frac{\nabla^i\Omega\nabla^j\Omega}{\Omega^2}}\right].
\end{eqnarray}
The spatial Ricci and scalar curvatures are
\begin{eqnarray}
%
%
 \3\1{R}_{ij}({\gamma_{ij}}) 
	&=& - \left[ \frac{4}{3} \frac{a_\pm'\Omega}{a_\pm} 
		-\frac{2}{3}\bar F \right] 
		\Oll h^\pm _{ij} \nonumber \\
	&&+\frac{2}{3}\left[-2\frac{a_\pm''\Omega^2}{a_\pm}
	+\left(\frac{a_\pm '\Omega}{a_\pm}\right)^2   
	+{\bar F}'\Omega-\bar F +\frac{a_\pm'\Omega}{a_\pm}\bar F -{\bar F}^2\right]
		\OlOl h^\pm _{ij} \nonumber \\
	&&-\left[\frac{a_\pm '\Omega}{a_\pm} 
	-2\bar F \right]\overline{\Oij} \nonumber \\
 	&&+\left[-\frac{a_\pm ''\Omega^2}{a_\pm} 
		+2\left(\frac{a_\pm '\Omega}{a_\pm}\right)^2
		+2{\bar F}'\Omega-2\bar F -\frac{a_\pm '\Omega}{a_\pm}\bar F 
		+\frac{5{\bar F}^2}{2} \right]\overline{\OiOj} , \nonumber \\
%
%
 \3\1{R_i^{~j}}({\gamma_{ij}}) 
	&=& -\frac{1}{a_\pm^2}\left\{ \frac{2}{3}\left[ 
  		\pm F-\bar F +2\frac{a_\pm '\Omega}{a_\pm}\right]
		\Oll \delta^j_i \right. \nonumber \\
 	&&+\frac{2}{3}\left[\pm G+\bar F -{\bar F}'\Omega
		\pm 2{\bar F}^2-{\bar F}^2
 	+\Omega^2\left(	2\frac{a_\pm ''}{a_\pm} 
		-\frac{a_\pm '^2}{a_\pm^2}\right)\right] \OlOl\delta^j_i \nonumber \\
	&&+\left[\frac{a_\pm '\Omega}{a_\pm}\pm 2\bar F-2\bar F \right]
	\overline{\frac{\nabla_i\nabla^j\Omega}{\Omega}} \nonumber \\
	&&+\left.\left[ \pm 2\bar G +2\bar F 
		-2{\bar F}'\Omega-(\pm 2+\frac{1}{2}){\bar F}^2
 	-2\left(\frac{a_\pm '\Omega}{a_\pm}\right)^2 
	+\frac{a_\pm ''\Omega^2}{a_\pm}\right]
 	\overline{\frac{\nabla_i\Omega\nabla^j\Omega}{\Omega^2}}\right\},
  \nonumber  \\
 \3\1{R}({\gamma_{ij}})
	&=&-\frac{1}{a_\pm^2}\left\{2\left[\pm F-\bar F 
	+2\frac{a_\pm '\Omega}{a_\pm}\right]
		\Oll \right. \nonumber \\
	&&\left.+2\left[\pm G +\bar F - {\bar F}'\Omega 
		\pm 2 {\bar F}^2 - {\bar F}^2
	+\Omega^2 \left(2\frac{a_\pm ''}{a_\pm} 
	-\frac{a_\pm '^2}{a_\pm^2}\right)\right]
	\OlOl  \right\} ,
\label{eq:scalarR}
\end{eqnarray}
where $a_\pm'\equiv \partial a_\pm/\partial\Omega$.

\end{appendix}

\pagebreak

\begin{figure}[htbp]
\begin{center}
  \leavevmode
  \epsfbox{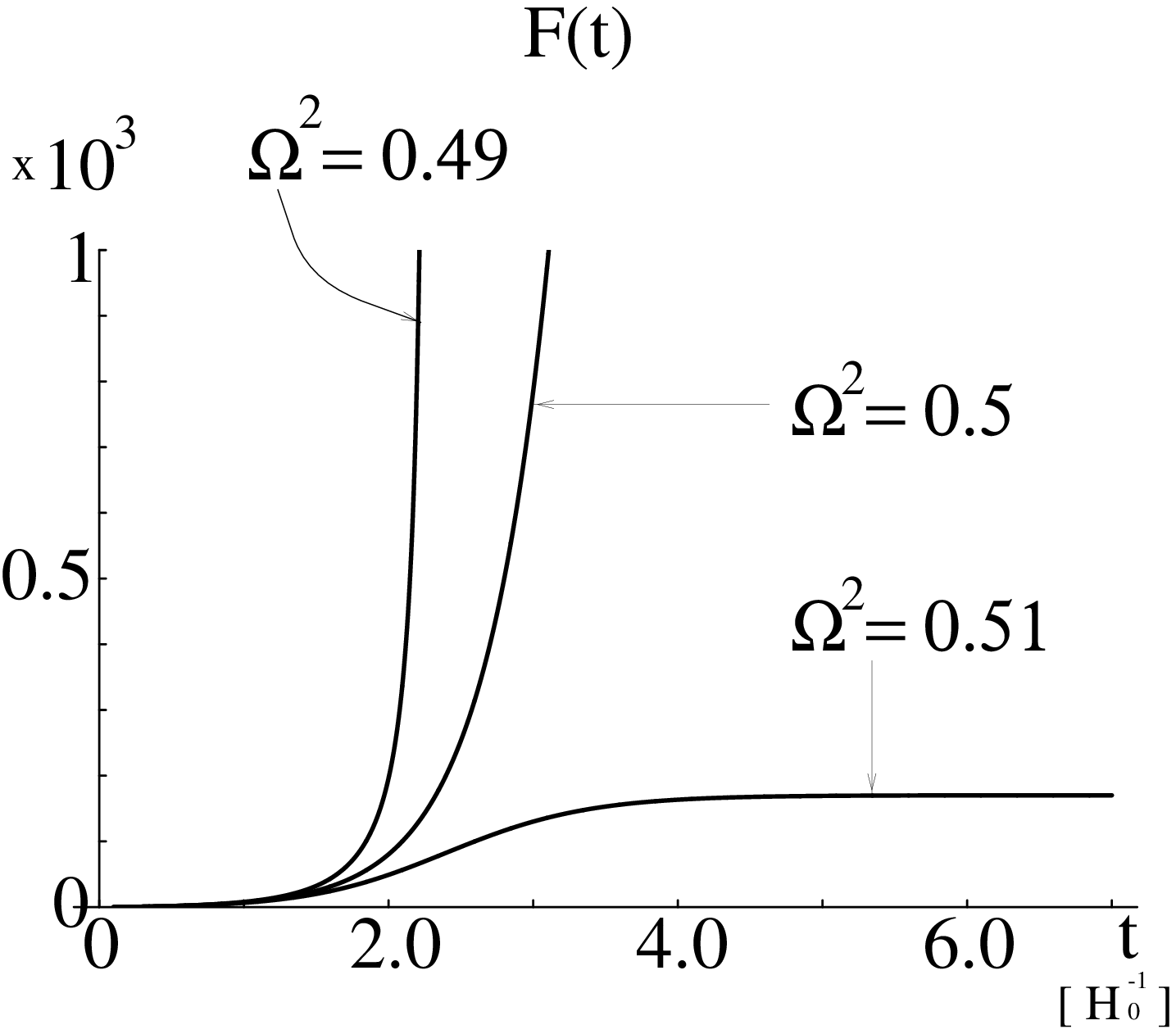}
\caption{}
\label{fig:F}
\end{center}
\end{figure}

\noindent
{\Large\bf Figure Captions}
\vskip 5mm
\noindent 
Fig.\ref{fig:F}.

A typical example of the time evolution of $F(t)$.

\begin{figure}[htbp]
\begin{center}
  \leavevmode
  \epsfbox{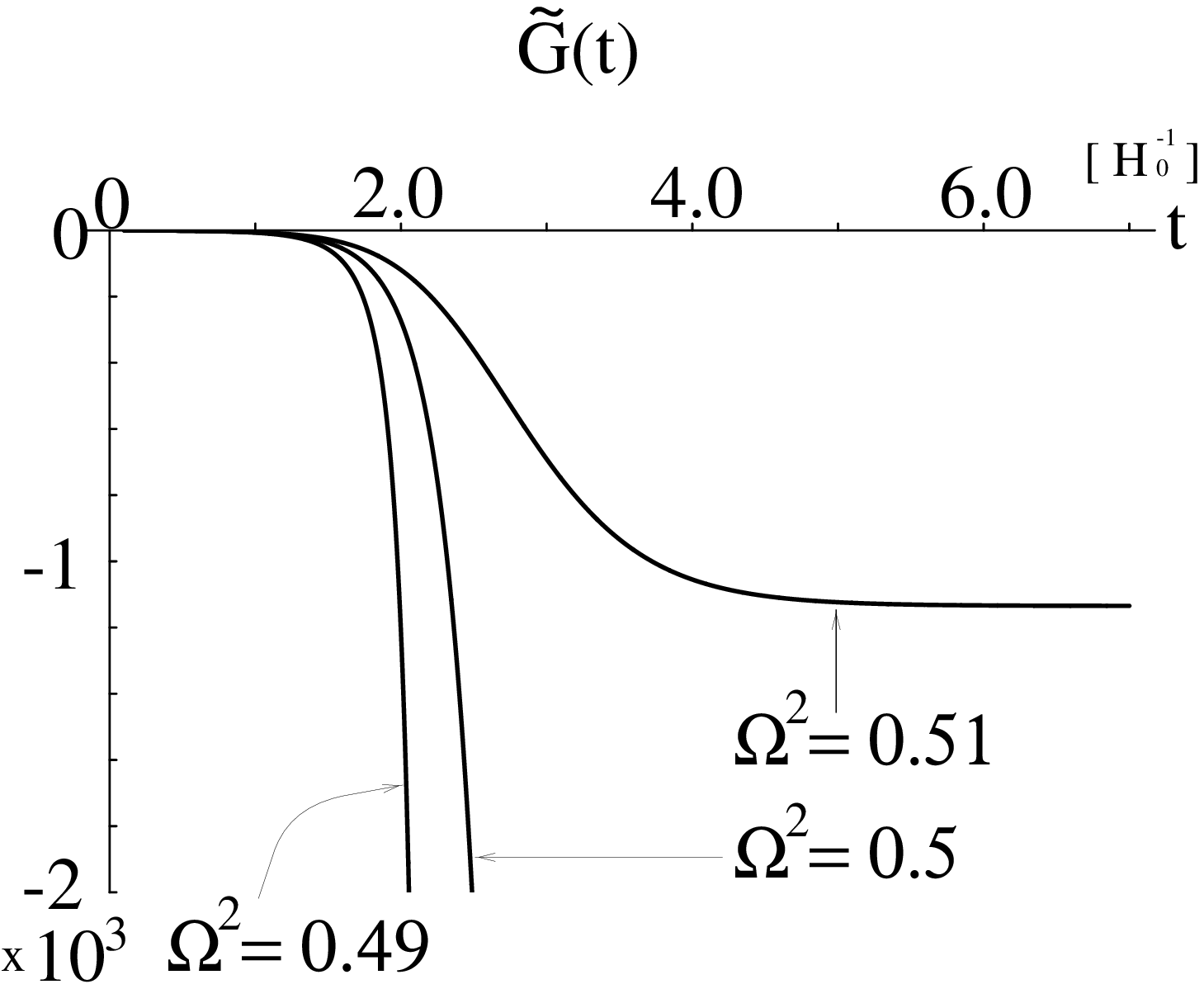}
\caption{}
\label{fig:G}
\end{center}
\end{figure}

\noindent 
Fig.\ref{fig:G}.

A typical example of the time evolution of $\tilde G(t)$.

\begin{figure}[htbp]
\begin{center}
  \leavevmode
  \epsfbox{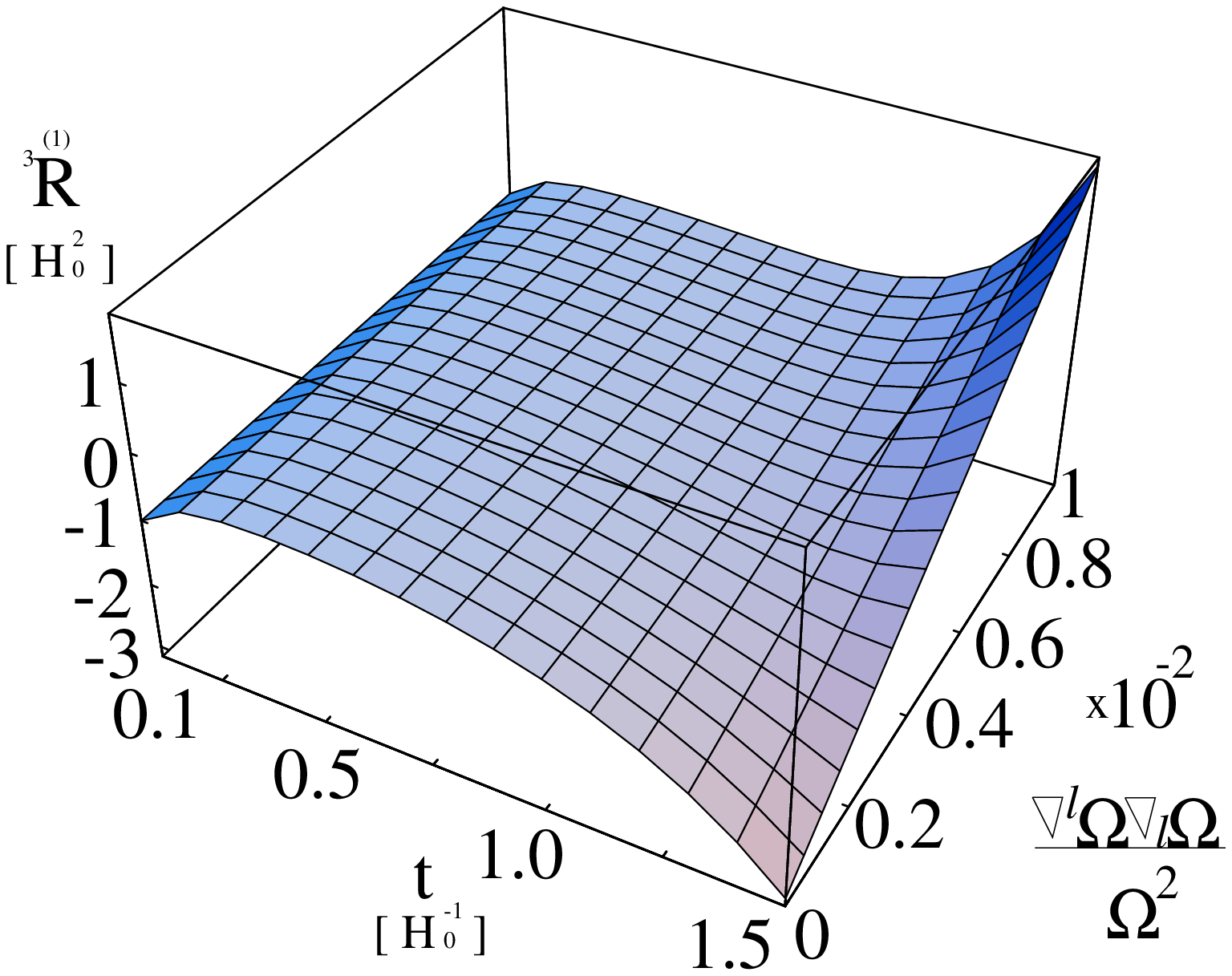}
\caption{}
\label{fig:R}
\end{center}
\end{figure}

\noindent 
Fig.\ref{fig:R}.

Evolution of the spatial curvature which has the same initial 
value but different $\nabla_l\Omega\nabla^l\Omega/\Omega^2$. 

\begin{figure}
\begin{center}
  \leavevmode
  \epsfbox{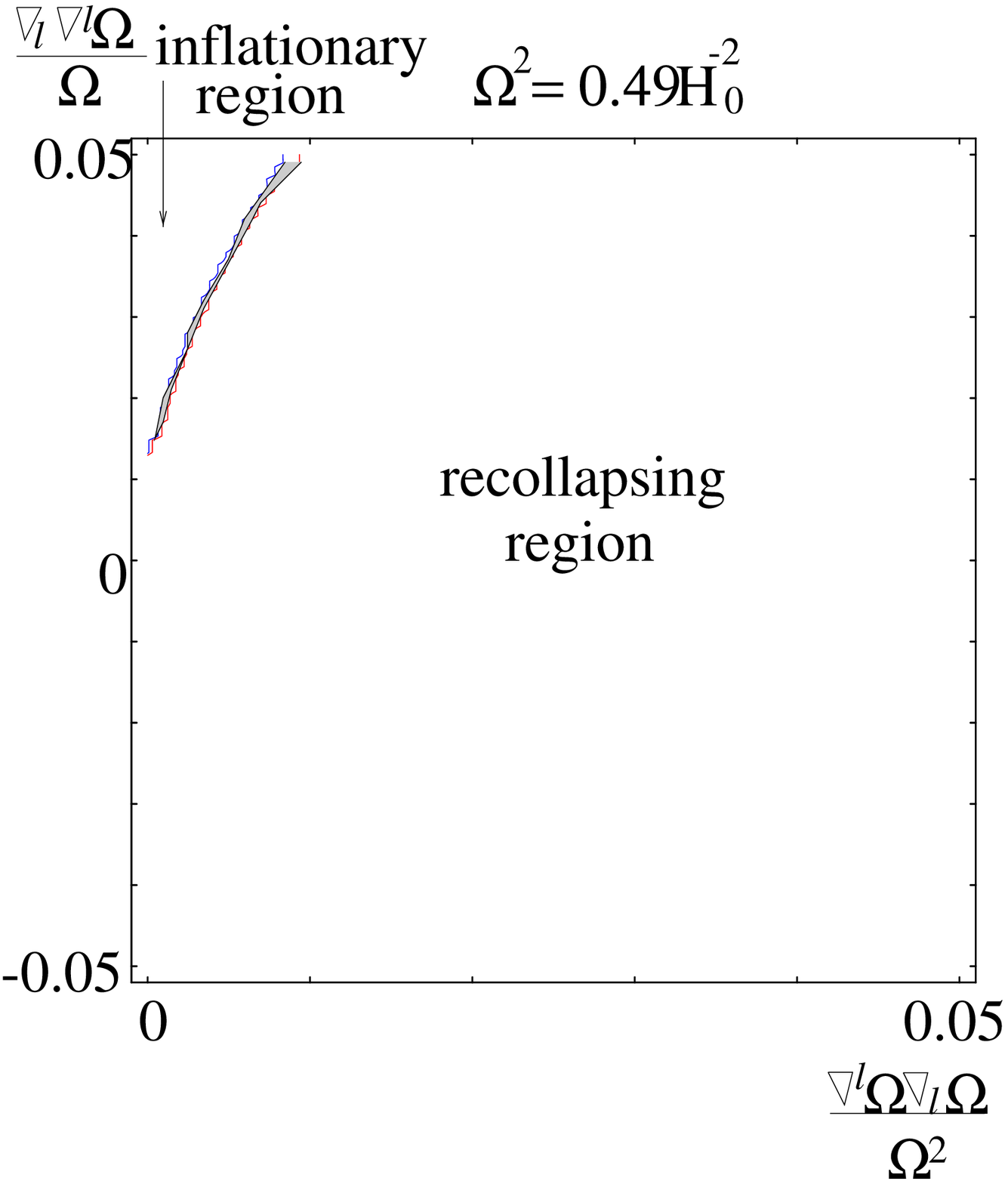}
\caption{}
\label{fig:cr49}
\end{center}
\end{figure}

\begin{figure}[htbp]
\begin{center}
  \leavevmode
  \epsfbox{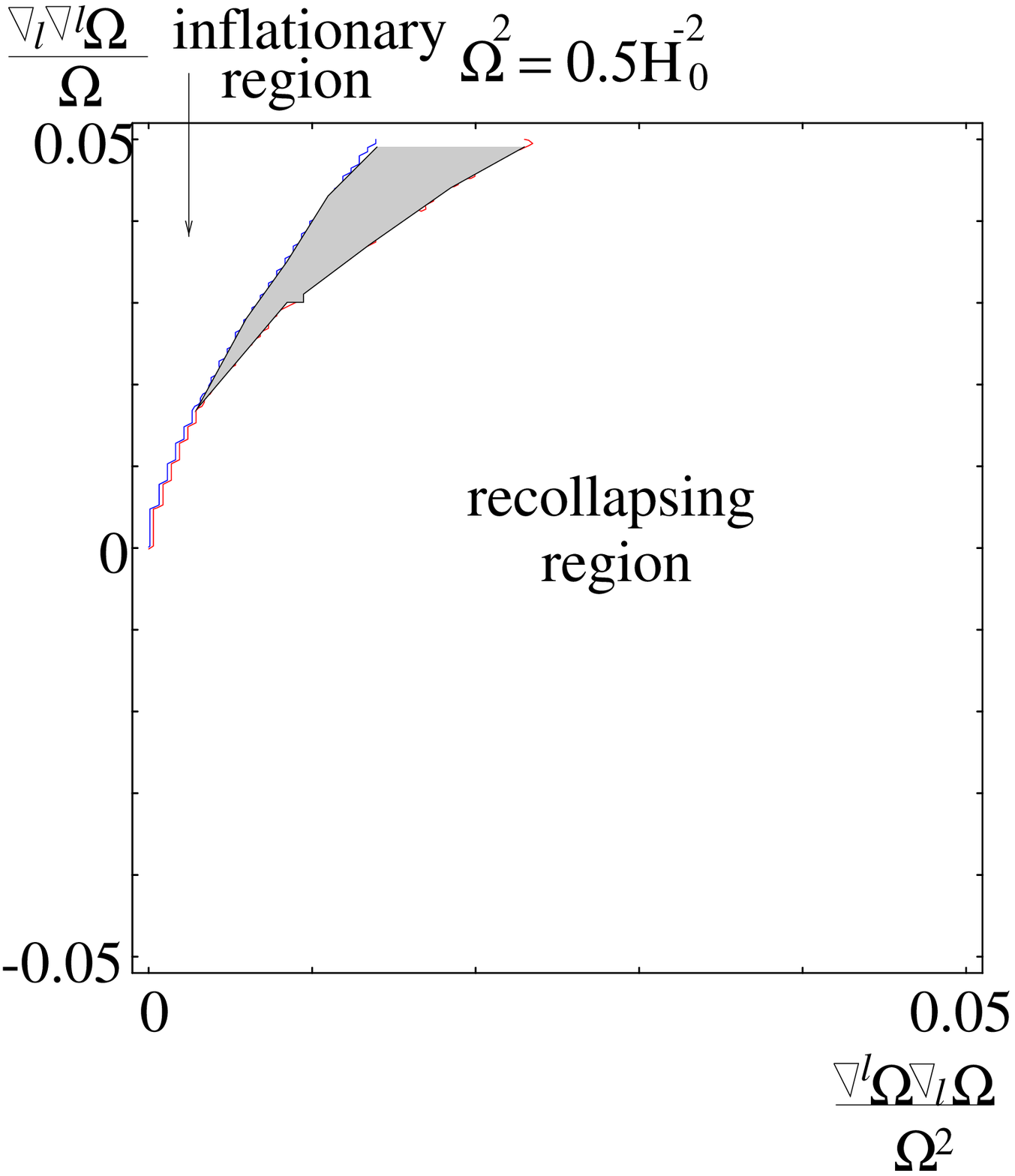}
\caption{}
\label{fig:cr5}
\end{center}
\end{figure}

\begin{figure}[htbp]
\begin{center}
  \leavevmode
  \epsfbox{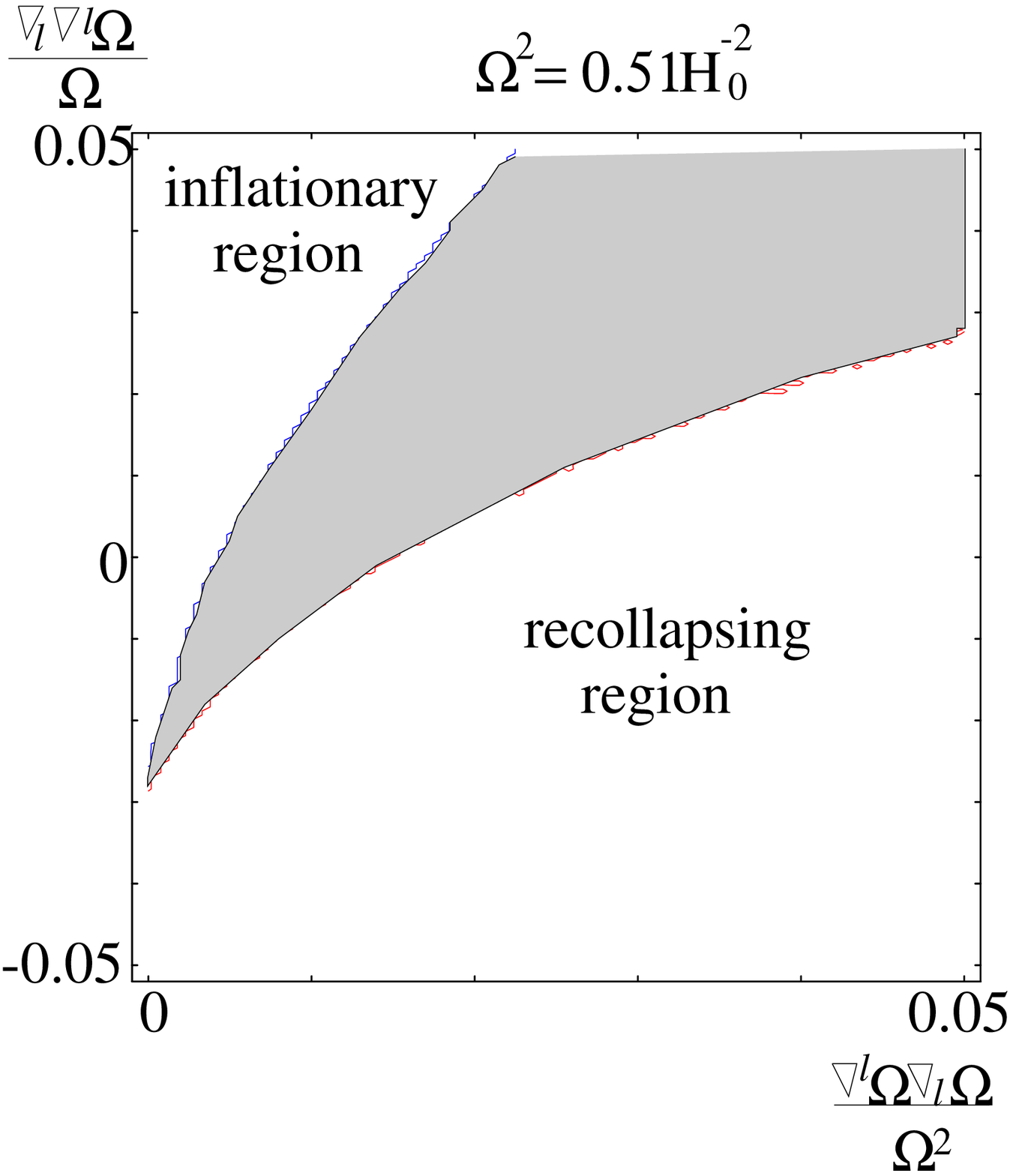}
\caption{}
\label{fig:cr51}
\end{center}
\end{figure}

\noindent 
Figs.\ref{fig:cr49} - \ref{fig:cr51}.

An inflationary region and a recollapsing region are shown 
in the three dimensional parameter space 
$(\Omega, \nabla_l\nabla^l\Omega/\Omega, \nabla_l\Omega\nabla^l\Omega/\Omega^2)$. 
A meshed region is a region where the approximation breaks down.




\newpage

\newpage

\newpage

\newpage

\end{document}